\documentclass{aa}  

\usepackage{graphicx}
\usepackage{txfonts}
\begin{document}

   \title{Water masers in Compton-thick AGN}

   \subtitle{II. The high detection rate and EVN observations of IRAS~15480-0344}

   \author{P. Castangia
          \inst{1}
          \and
          G. Surcis\inst{1}%\fnmsep\thanks{}
          \and
          A. Tarchi\inst{1}
          \and
          A. Caccianiga\inst{2}
          \and
          P. Severgnini\inst{2}
          \and
          R. Della Ceca\inst{2}
          }

   \institute{INAF-Osservatorio Astronomico di Cagliari, Via della Scienza 5, 09047, Selargius (CA), Italy\\
              \email{pcastang@oa-cagliari.inaf.it}
         \and
             INAF-Osservatorio Astronomico di Brera, Via Brera 28, 20121, Milan, Italy\\
             %\email{}
             %\thanks{}
             }

   \date{}

% \abstract{}{}{}{}{} 
% 5 {} token are mandatory
 
  \abstract
   {}
   {Investigations of H$_2$O maser galaxies at X-ray energies reveal that most AGN associated with water masers are characterized by high levels of absorption. With the aim of finding new maser sources for possible interferometric follow-ups, we have searched for water maser emission in a well-defined sample of heavily absorbed AGN ($N_{\rm H} > 10^{23}$\,cm$^{-2}$), including Compton-thick (CT) sources.}
   {All the galaxies in the sample were already searched for 22\,GHz water maser emission in previous surveys. With the goal of providing a detection or a stringent upper limit on the H$_2$O luminosity, we re-observed some of the non-detected sources with the Green Bank Telescope. A new luminous H$_2$O maser ($L_{\rm H2O} \sim 200\,$L$_\odot$) was detected in the mid-IR-bright Seyfert~2 galaxy IRAS~15480-0344 and then followed-up with the Very Long Baseline Array. In order to shed light on the origin of the maser (jet/outflow vs. disc), we recently observed the radio continuum emission in IRAS~15480-0344 with the European VLBI network (EVN) at 1.7 and 5.0\,GHz.}
   {With the newly discovered megamaser in IRAS~15480-0344 revealing a narrow ($\sim$0.6\,km\,s$^{-1}$) and a broad ($\sim$90\,km\,s$^{-1}$) component \citep{castangia2016}, the maser detection rate of the CT AGN sample is 50\% (18/36), which is one of the highest ever found in maser surveys. The EVN maps show two bright sources (labeled SW and NE) in the nuclear region of IRAS~15480-0344, which we interpret as jet knots tracing regions where the radio plasma impacts dense molecular clouds. The narrow maser feature is approximately at the centre of the imaginary line connecting the two continuum sources, likely pinpointing the core, and might be associated with the accretion disc or a nuclear outflow. The location of the broad maser feature, instead, coincides with source NE, suggesting that the maser emission might be produced by a jet-cloud interaction, as it was proposed for NGC~1068 and Mrk~348.}
   {} 


   \maketitle
%
%________________________________________________________________

\section{Introduction} % da cambiare
A key ingredient in the Unified Model of active galactic nuclei (AGN) is the dusty toroidal structure, known as the ``torus'' \citep{antonucci93,urry95}, surrounding the accreting supermassive black hole (SMBH). This structure is supposed to block the direct emission produced in the accretion disc by scattering it and re-emitting it in the infrared (IR). Studies of the IR and X-ray emissions in AGN over the past 10-15 years have provided relevant information on the obscuring matter in the vicinity of SMBHs. As a consequence, the classical, uniform, dusty  torus, seen as an isolated entity, has now been replaced by a clumpy structure that is connected physically and dynamically with the host galaxy via gas inflows/outflows \citep[for a recent review see][]{ramosalmeida2017}. The torus radiates most of its energy at mid-IR wavelengths. Mid-IR interferometry has revealed that the emitting dust is concentrated on scales of 0.1--10\,pc and, in most cases, can be modeled with two nuclear components, instead of a single disc/toroidal structure \citep{burtscher2013}. In some sources one of these components is elongated in the polar direction and has been interpreted as an outflowing dusty wind driven by radiation pressure \citep[e.\,g.,][]{hoenig2012}. 

One of the most recent models for the IR emission in AGN, is based on the premise that the dusty gas around AGN consists of an inflowing disc and an outflowing wind. The disc gives rise to the 3--5\,$\mu$m near-IR component, while the wind produces the mid-IR emission \citep{hoenig2017}. X-ray absorption variability studies, instead, have demonstrated that the obscuring material is not homogeneous but clumpy and dynamic (with clouds being created and dissipated continuously) and located at various spatial scales, from the broad line region (BLR) to the torus \citep[][and references therein]{ramosalmeida2017}. Despite the great progress made, however, the exact geometry of the absorbing matter (e.\,g., is the torus geometrically thick? Is the polar elongation always present?) and its dynamical origin (are the BLR and the torus produced by accretion disc winds?) are not yet fully understood. The study of the physical properties, the structure, and the kinematics of the gas surrounding SMBHs, is fundamental to answer these open questions and to build detailed models of AGN. Furthermore, it may be relevant also to shed light on the impact of nuclear activity on galaxy evolution. Indeed, AGN-driven outflows may influence galaxy evolution by regulating star formation \citep[e.\,g.,][]{dimatteo05}. AGN feedback may be caused by radiative winds from the accretion discs or by outflows produced by radio jets as they interact with the interstellar medium (ISM) of the host galaxy \citep{wylezalek2018}. Constraining the main driving mechanism of outflows in AGN is essential to understand their effect on galaxy evolution. 

Due to the small dimensions of the torus, current IR and X-ray instruments are not able to resolve it and information on its structure have to be inferred by modelling the emitted radiation. The radio emission from luminous H$_2$O masers (the so-called ``megamasers'') constitutes the only way to directly map the gas at sub-parsec distance from the SMBH \citep[for recent reviews see][]{lo05, greenhill07,tarchi2012,henkel2018,braatz2018}. The high brightness temperature and small size of the maser spots make them perfect targets for Very Long Baseline Interferometry (VLBI) observations, through which angular resolutions of the order of 0.1\,mas can be reached. Interferometric and single-dish monitoring studies of water maser sources allow us to determine accretion disc geometry and to estimate the enclosed dynamical masses \citep[e.g.][]{kuo2011,gao2017,zhao2018}. In addition, radio continuum observations of disc-maser galaxies have been recently used to test some aspects of the AGN paradigm, i.\,e. the alignment bewteen the radio jet and the rotation axis of the accretion disc \citep[][and references therein]{kamali2019}. H$_2$O masers may also trace nuclear outflows in the form of jets or winds. Jet-maser observations can provide estimates of the shock speeds and densities of radio jets, improving our understanding of the jet-ISM interaction \citep{gallimore01,peck03}. Water maser observations in Circinus \citep{greenhill03} and NGC~3079 \citep{kondratko05}, instead, seem to have resolved individual outflowing torus clouds at $<$1\,pc from the nuclear engine. Proper motion measurements and comparison of these outflow-masers with their disc counterpart have the potential to probe the structure and kinematics of the torus molecular clouds \citep{nenkova08}. Therefore, each megamaser source provides a wealth of information on the (sub-)parsec-scale environment around AGN, making the discovery of new sources and their interferometric follow-up extremely important for AGN studies.
 
\vspace{0.25cm}
\noindent {\bf The sample and the new H$_2$O maser in IRAS~15480-0344}\\
We searched for 22\,GHz water maser emission in a well defined sample of 36 heavily absorbed AGN ($N_{\rm H} > 10^{23}$\,cm$^{-2}$), including Compton-thick (CT) sources (Table~\ref{table:sample}), selected in the local Universe through a combination of mid-IR (\textit{IRAS}) and X-ray (\textit{XMM-Newton}) data \citep[for details, see][]{severgnini2012}. All the galaxies in the sample were already observed at 22\,GHz in previous surveys, and water maser emission was detected in 17/36 of them. With the goal of providing a detection or a stringent upper limit on the H$_2$O luminosity, we re-observed some of the non-detected sources with the Green Bank Telescope (GBT). These new observations led to the discovery of a new luminous H$_2$O maser ($L_{\rm H2O} \sim 200\,$L$_\odot$) in the mid-IR-bright Seyfert~2 galaxy IRAS~15480-0344 \citep[hereafter IRAS15480;][]{castangia2016} and improved upper limits on 4 objects. Based on the single-dish profile, the variability of the maser emission, and the location of the maser spots inferred from VLBI observation, we interpreted the line emission in IRAS15480 as the result of a jet/outflow interaction \citep{castangia2016}. However, an alternative scenario in which the whole maser emission is produced in a slowly rotating accretion disc, could not be ruled out by observational data. Indeed, it was not possible to associate the position of the line emission with other sources of activity in the nuclear region of the galaxy, due to the lack of images with an angular resolution comparable to that of our Very Long Baseline Array (VLBA) spectral line data. To overcome this limitation, we observed the nuclear radio continuum in IRAS15480 with the European VLBI Network (EVN). Here we present the overall results of the survey, leaving a detailed statistical analysis of the CT AGN sample to a forthcoming paper, and report the outcome of the EVN observations of IRAS15480. Section~\ref{sect:obs} describes the details of the GBT and EVN observations and data reduction. The results are reported in Section~\ref{sect:results}. We discuss the high detection rate obtained in the survey in Section~\ref{sect:discussion}. In this Section, we also examine the possible scenarios for the radio continuum emission in the nucleus of IRAS15480 and discuss the origin of the maser in light of the new EVN data. We draw our conclusions in Section~\ref{sect:conclusions}. Throughout the paper we adopt a cosmology with $\Omega_{\rm M} =0.3$, $\Omega_{\rm \Lambda} =0.7$ and $H_0 = 70$\,km\,s$^{-1}$\,Mpc$^{-1}$. The quoted velocities are calculated using the optical velocity definition in the heliocentric frame. 

\begin{table*}
\caption{Compton-thick AGN of the 2XMM--IRAS sample}
\label{table:sample}      
\centering          
\begin{tabular}{cccrrcl}     % 6 columns 
\hline\hline
IRAS name & Other name & Redshift  & $rms$ & $L_{\rm H2O}$ & Maser nature\tablefootmark{a} & References\tablefootmark{b} \\ 
          &            &           & (mJy) & (L$_{\odot}$) &              &           \\ 
\hline                    
${\bf 00085-1223}$\tablefootmark{c}  & {\bf NGC 17}    &  0.0196  &     & {\bf 60}  & ?  & gre09 \\       
$00387+2513$        &  NGC 214        &  0.0151  & 4   & $<$15     &    & bra08 \\    
$01091-3820$        &  NGC 424        &  0.0118  & 12  & $<$24     &    & kon06 \\       
${\bf 01306+3524}$  & {\bf NGC 591}   &  0.0152  &     & {\bf 38}  & D  & pes15 \\       
$01413+0205$        &  Mrk 573	      &  0.0172  & 3   & $<$10     &    & bra04 \\       
${\bf 02401-0013}$  & {\bf NGC 1068}  &  0.0038  &     & {\bf 200} & D+J & ben09 \\      
${\bf 03012-0117}$  & {\bf NGC 1194}  &  0.0136  &     & {\bf 131} & D   & pes15 \\      
$03106-0254$        &  2MFGC 2636     &  0.0272  & 11  & $<$120    &     & kon06 \\       
${\bf 03222-0313}$  &  {\bf NGC 1320} &  0.0089  &     & {\bf 30}  & D   & gre09 \\       
$03317-3618$        &  NGC 1365       &  0.0055  & 3   & $<$1      &     & bra08 \\       
${\bf 03348-3609}$  &  {\bf NGC 1386} &  0.0029  &     & {\bf 120} & D+? & bra96 \\  
$04507+0358$        &  CGCG 420-015   &  0.0294  & 14  & $<$180    &     & kon06 \\
$05093-3427$        &  ESO 362-8      &  0.0157  & 18  & $<$70     &     & kon06 \\      
${\bf 06097+7103}$  & {\bf Mrk 3}     &  0.0135  &     & {\bf 5}   & ?   & MCP  \\       
${\bf 06456+6054}$  & {\bf NGC 2273}  &  0.0061  &     & {\bf 37}  &  D  & pes15 \\      
${\bf 07379+6517}$  & {\bf Mrk 78}    &  0.0372  &     & {\bf 104} &  D  & pes15 \\
$08043+3908$        &  Mrk 622	      &  0.0232  & 3   & $<$24     &     & bra04 \\
${\bf 09320+6134}$  & {\bf UGC 05101} &  0.0394  &     & {\bf 1762} & ?  & zha06 \\ 
${\bf 09585+5555}$  & {\bf NGC 3079}  &  0.0037  &     & {\bf 230} & D+O & MCP  \\
$11538+5524$        &  NGC 3982       &  0.0037  & 3   & $<$1      &     & bra04  \\	
$12540+5708$        &  Mrk 231	      &  0.0422  & 7   & $<$190    &     & zha06  \\  
${\bf 13044-2324}$  & {\bf NGC 4968}  &  0.0099  &     & {\bf 53}  & ?   & MCP  \\  
${\bf 13277+4727}$  & {\bf  M51}      &  0.0020  &     & {\bf 2}   & ?   & ho87 \\  
${\bf 13362+4831}$  & {\bf Mrk 266}   &  0.0279  &     & {\bf 30}  & ?   & bra04  \\  
13428+5608          &  Mrk 273        &  0.0378  & 2   & $<$42     &     & this work \\  
15295+2414          &      3C 321     &  0.0961  & 0.9 & $<$120    &     & this work \\
$15327+2340$        &  Arp220         &  0.0181  & 10  & $<$50     &     & hen05   \\
${\bf 15480-0344}$  & {\bf 2MASXJ15504152-0353175} & 0.0303 & & {\bf 200} & J+? & cas16 \\
${\bf 16504+0228}$  & {\bf NGC 6240}  &  0.0245  &     & {\bf 40}  & ?   & MCP  \\   
$18429-6312$        &  IC 4769        &  0.0151  & 13  & $<$40     &     & kon06  \\  
$19254-7245$        &  AM 1925-724    &  0.0617  & 125 & $<$7030   &     & gre02 \\  
${\bf 20305-0211}$  & {\bf NGC 6926}  &  0.0196  &     & {\bf 410} &  ?  & MCP  \\
22045+0959          &      NGC 7212   &  0.0267  & 2   & $<$20     &     & this work \\
${\bf 23024+1203}$  & {\bf NGC 7479}  &  0.0079  &     & {\bf 12}  &  ?  & MCP  \\
$23156-4238$        &  NGC 7582       &  0.0053  & 8   & $<$3      &     & sur09  \\
23254+0830          &   NGC 7674      &  0.0289  & 2   & $<$25     &     & this work  \\
%Gli rms sono calcolati su canali di 1.1-1.3 km/s
%Per le distanze ho usato cz/H0 con H0=70 km/s Mpc
\hline
\end{tabular}
\tablefoot{
\tablefoottext{a}{Origin of the maser emission: D -- disc; J -- jet; O -- outflow.}
\tablefoottext{b}{References for the maser luminosity or the 1$\sigma$ r.m.s: ben09 -- \citet{bennert09}; bra96 -- \citet{braatz96}; bra04 -- \citet{braatz04}; bra08 -- \citet{braatz08}; cas16 -- \citet{castangia2016}; gre02 -- \citet{greenhill02}; gre09 -- \citet{greenhill09}; hen05 -- \citet{henkel05}; ho87 -- \citet{ho87}; kon06 -- \citet{kondratko06}; MCP -- Megamaser Cosmology Project; pes15 -- \citet{pesce2015}; sur09 -- \citet{surcis09}; zha06 -- \citet{zhang06}.}
\tablefoottext{c}{The sources in bold are the galaxies known to host water maser emission.}
}
\end{table*}

\section{Observations and data reduction}\label{sect:obs}

\subsection{GBT observations}\label{sect:gbtobs}
Observations were conducted with the GBT between March 9, 2012 and March 4, 2013. We employed two of the seven beams of the K-band focal plane array (KFPA) receiver in total power nod mode and configured the spectrometer with two 200\,MHz IFs, each with 8192 channels. This observational setup yielded a channel spacing of 24\,kHz, corresponding to $\sim$0.3\,km\,s$^{-1}$ at the frequency of 22\,GHz. The data were reduced and analysed using the GBTIDL package\footnote{http://gbtidl.nrao.edu/}. More details on the observations and data reduction are described in \citet{castangia2016}. In Table~\ref{table:gbtobs} we report, for each galaxy observed with the GBT but not detected, the observation date, the inspected velocity range, and the 1$\sigma$ root mean square (rms) noise of the spectrum referred to a $\sim$1.4\,km\,s$^{-1}$ wide channel. 

\begin{table*}
\caption{Galaxies re-observed with the GBT but not detected}             
\label{table:gbtobs}      
\centering          
\begin{tabular}{c c c c l l}     % 6 columns 
\hline\hline       
IRAS name & Other name  & $V_{\rm sys}$\tablefootmark{a} & Obs. date & $\Delta V$    & $rms$\tablefootmark{b} \\
          &             & (km\,s$^{-1}$)                &           & (km\,s$^{-1}$) & (mJy) \\
\hline
$13428+5608$  & Mrk 273 &  11326         & 12-04-2012 & 9871--15436  & 2     \\  
$15295+2414$  & 3C 321  &  28810         & 09-03-2012 & 27184--33394 & 1.4   \\ 
              &         &                & 07-04-2012 &              & 2.4   \\ 
              &         &                & 12-05-2012 &              & 2.3    \\
              &         &                & 02-02-2013 &              & 1.6    \\
              &         &                & average    &              & 0.9    \\
$22045+0959$  & NGC 7212 &  7983         & 04-03-2013 & 6568--12017  & 2      \\
$23254+0830$  & NGC 7674 &  8671         & 07-04-2012 & 7246--12722  & 2      \\
%Gli rms sono calcolati su canali di 0.3x4=1.2 km/s
\hline                  
\end{tabular}
\tablefoot{
\tablefoottext{a}{Systemic velocity of the observed galaxies, calculated using the optical definition in the heliocentric frame.}
\tablefoottext{b}{Inspected velocity range.}
\tablefoottext{c}{The rms refers to a $\sim$1.4\,km\,s$^{-1}$ wide channel.}
}
\end{table*}

\subsection{EVN observations and imaging}\label{sect:evnobs}
We observed the nucleus of IRAS15480 with the EVN\footnote{The European VLBI Network is a joint facility of independent European, African, Asian, and North American radio astronomy institutes.} at 1.7 and 5\,GHz, between February and March, 2015, employing a sensitive array of EVN antennas (see Table~\ref{table:vlbiobs}). The data were recorded at 1024\,Mbps, with 8$\times$16\,MHz IFs and dual circular polarization. Cross-correlation of the data was performed using the EVN software correlator \citep[SFXC;][]{keimpema2015} at the Joint Institute for VLBI ERIC (JIVE), using 32 channels per IF and polarization. We observed in phase-referencing mode, to correct phase variation caused by the atmosphere and estimate absolute positions. We used J1555-0326 as a phase calibrator. The strong compact sources 3C345, J1751+0939, and J2005+7752, were observed as fringe finders. The total observing time was six hours for each frequency band.

\begin{table*}
\caption{EVN observation and map details for IRAS15480.}
\label{table:vlbiobs}      
\centering          
\begin{tabular}{lllrrl}     % 6 columns 
\hline\hline
Date & Frequency & Partecipating stations\tablefootmark{a} & Synth. beam & P.A.         & $rms$ \\ 
     & (GHz)     &                                         &   (mas$\times$mas)     & ($\degr$)    & (mJy\,beam$^{-1}$) \\ 
\hline      
2015-Feb-28   & 1.7 & Ef, Wb, On, Jb1, (Nt), Sv, (Mc), Tr, Zc, Ur, Bd, Sr, (Sh) & 14$\times$8   &  4           & 0.1 \\
2015-Mar-11   & 5.0 & Ef, Wb, On, Jb2, Nt, Sv, Mc, Tr, Zc, (Ur), Bd, Ys, Sh     & 5$\times$3    & -3           & 0.04 \\
\hline
\end{tabular}
\tablefoot{
\tablefoottext{a}{Station codes are as follows. Bd: Badary: Ef: Effelsberg; Jb1: Jodrell Bank (Lovell telescope); Jb2: Jodrell Bank (Mk2); Mc: Medicina; Nt: Noto; On: Onsala; Sh: Shangai; Sr: Sardinia Radio Telescope; Sv: Svetloe; Tr: Torun; Ur: Urumqi; Wb: Westerbork Synthesis Radio Telecope; Ys: Yebes; Zc: Zelenchukskaia. Telescopes in parentheses were scheduled but did not take part in the observations or did not produce good data due to technical problems.}
}
\end{table*}

We reduced and analysed the data utilizing the NRAO Astronomical Image Processing System (AIPS\footnote{http://www.aips.nrao.edu/}). Initial calibration, in particular amplitude calibration, was carried out a priori via the standard EVN pipeline. We then calibrated the bandpass shape using all the fringe finders as bandpass calibrators. Subsequently, we removed the instrumental delays by fitting the fringe patterns from J2005+7752. In order to solve for atmospheric phase variations, we fringe fitted the data from the phase reference source J1555-0326. Finally, we self-calibrated on J1555-0326 and then interpolated and applied the solutions to our target source IRAS15480. We followed the same calibration steps for both frequency bands, except that it was necessary to apply the ionospheric corrections as a first step of the calibration of the L-band dataset. The data were Fourier-transformed using uniform weighting and deconvolved using the CLEAN algorithm \citep{hoegbom74}. We mapped a field of 1$\times$1\,arcseconds$^2$ and 0.3$\times$0.3\,arcseconds$^2$ (corresponding to $\sim$600$\times$600\,pc$^2$ and $\sim$180$\times$180\,pc$^2$) at L- and C-band respectively, centered at the position of the nuclear radio continuum source visible in the VLA X-band image \citep[$\alpha_{2000}$=15$^{\rm h}$50$^{\rm m}$41$^{\rm s}$.498 and $\delta_{2000}$=$-$03\degr53\arcmin18\arcsec.05;][]{schmitt01}. Within the uncertainties, this source is coincident with the position of the optical nucleus \citep[$\alpha_{2000}$=15$^{\rm h}$50$^{\rm m}$41$^{\rm s}$.50 and $\delta_{2000}$=$-$03\degr53\arcmin18\arcsec.4;][]{klemola87}. The same position was chosen as the center of the 22\,GHz VLBA spectral line map \citep{castangia2016}. Details of the maps produced are reported in Table~\ref{table:vlbiobs}, where we indicate, for each map, the central frequency, the dimensions and position angles of the synthetized beams, and the rms.

\section{Results}\label{sect:results}

\subsection{Survey outcome}\label{sect:survey_res}
The 36 CT AGN of the \citet{severgnini2012} sample are listed in Table~\ref{table:sample}, where we also report, for each galaxy, the redshift, the 1$\sigma$ rms for 1.2--1.4\,km\,s$^{-1}$ channel, the isotropic line luminosity (or the upper limit), and the reference for the latter. For the galaxies re-observed with the GBT, we calculated the upper limits on the maser isotropic luminosity using the formula:
\begin{equation}
\label{eq:lum}
 L_{\rm H2O}[\rm L_{\odot}] = 0.023 \times S [\rm Jy] \times \Delta v [\rm km\,s^{-1}] \times D^2 [\rm Mpc^2],
\end{equation}
where $S$ and $\Delta v$ are the 5$\sigma$ rms limit in Jy and the line width in km\,s$^{-1}$, respectively. $D$ is the distance in Mpc. We assumed a width of ~$\sim$7\,km\,s$^{-1}$ for the maser lines (i.\,e., that the lines are detected above a 5\,$\sigma$ noise level in at least five 1.4\,km\,s$^{-1}$ channels).
When $L_{\rm H2O}$ or the upper limit were taken from the literature, we corrected them for $H_0 = 70$\,km\,s$^{-1}$\,Mpc$^{-1}$ if the authors used a different value of the Hubble's constant.  
For known maser sources, the table also contains a note indicating the origin of the line emission (disc, jet or outflow; Column~6), when this information was available in the literature. 
With the detection of IRAS15480, the maser detection rate of the entire sample becomes 50\%, which is one of the highest ever found in maser surveys.

\subsection{Nuclear continuum sources in IRAS15480}\label{sect:evn_res}
\begin{figure*}
\includegraphics[scale=0.45]{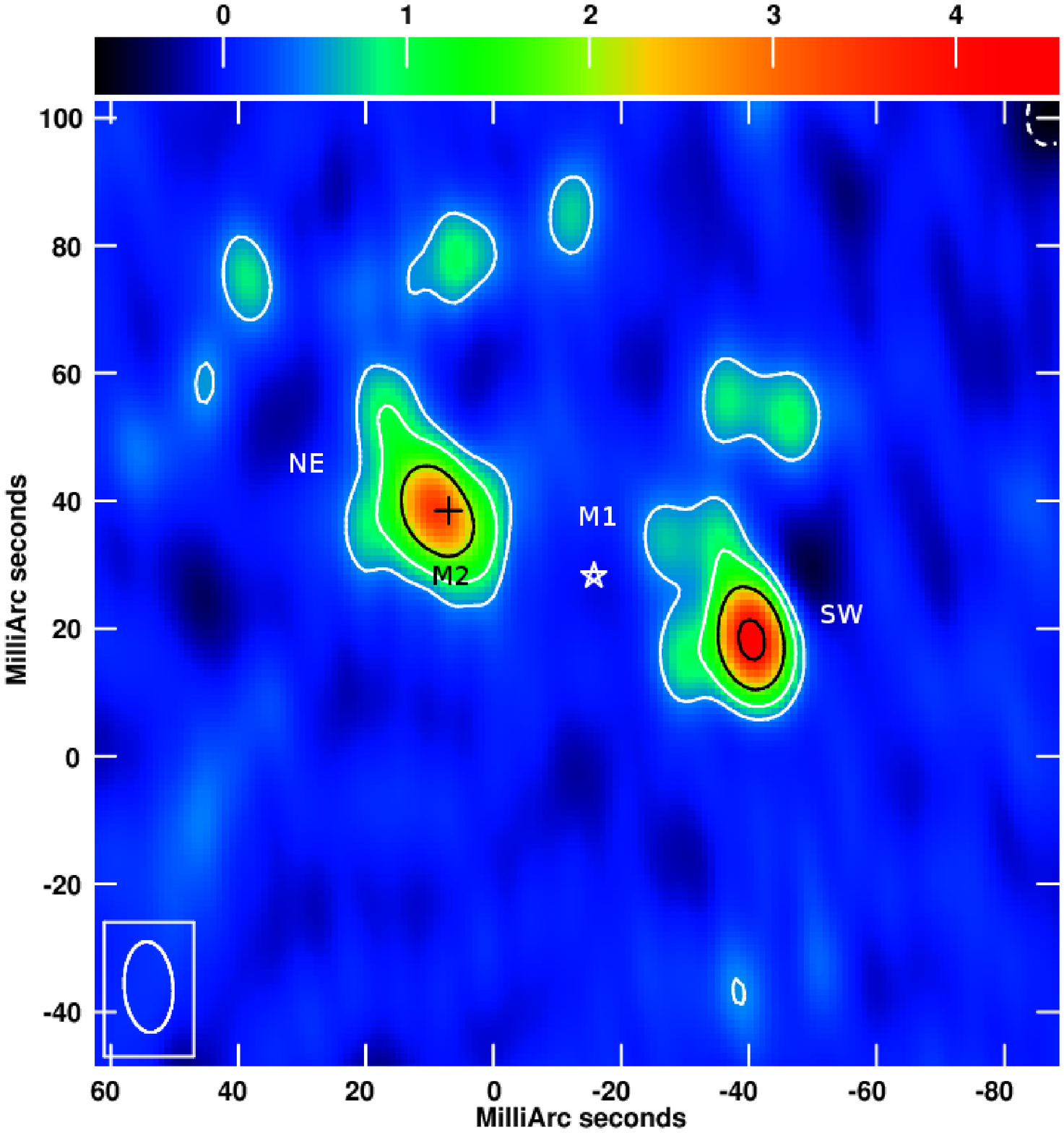}
\includegraphics[scale=0.45]{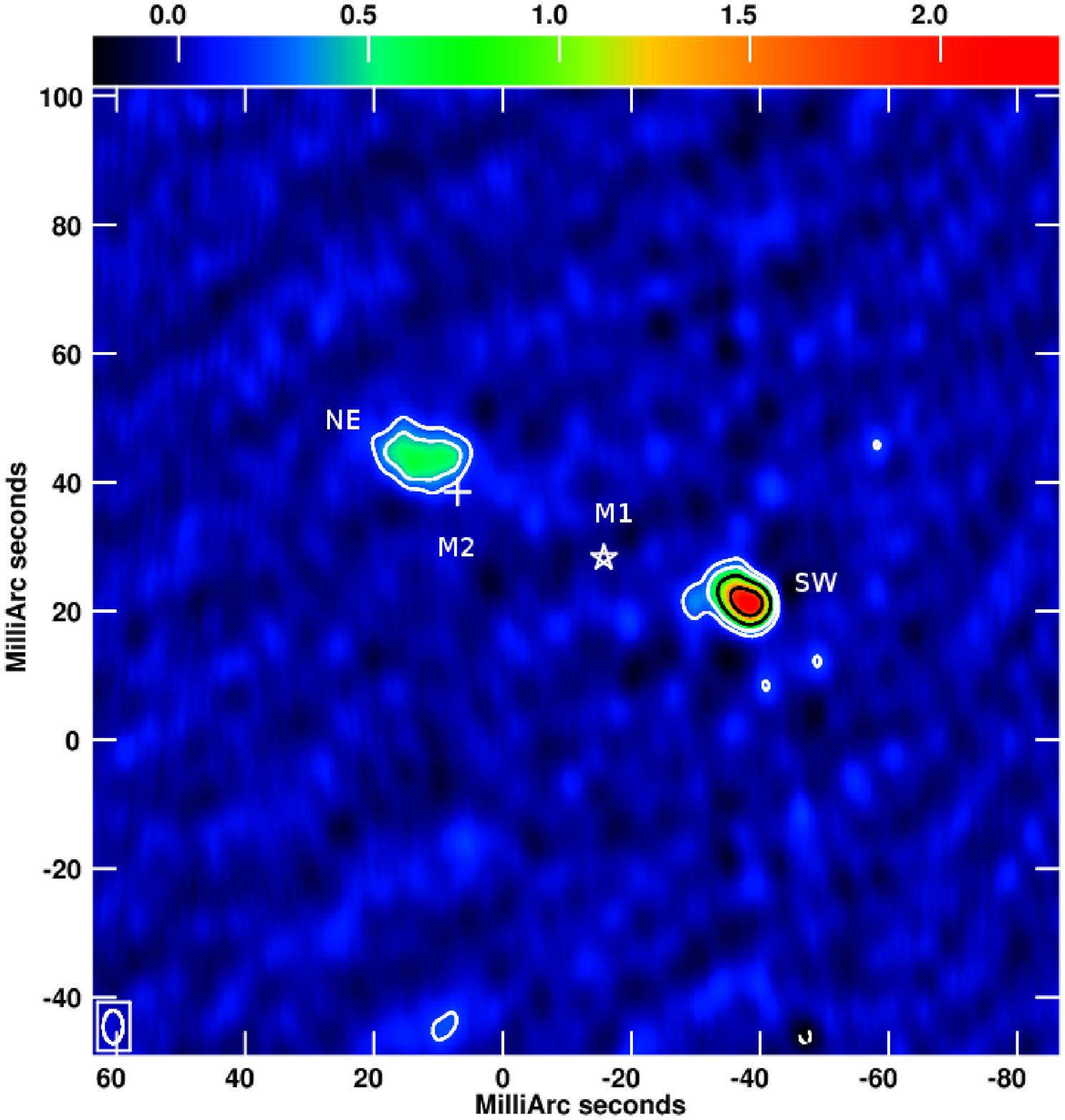}
\caption{EVN maps of the radio continuum emission in IRAS15480 at L-band (1.7\,GHz, {\it left}) and C-band (5\,GHz, {\it right}). Contour levels are (-1, 1, 2, 4, 8,...)$\times$ the 5$\sigma$ noise level, which is 0.5 and 0.2\,mJy/beam at 1.7 and 5\,GHz, respectively. The synthetized beams are shown in the lower-left corners. The positions of the water maser spots detected with the VLBA are also indicated: the star and the cross mark the location of the narrow (M1) and the broad blueshifted line emission (M2), respectively \citep[for details see][]{castangia2016}.}
\label{fig:cont+maser}
\end{figure*}

The EVN maps show two bright sources in the nucleus of IRAS15480 that have been detected at both 1.7\, and 5\,GHz (Fig.~\ref{fig:cont+maser}). 
The southwestern source (hereafter SW) is compact, being slightly resolved with dimensions of about $6 {\rm mas} \times 2 {\rm mas}$ ($4 {\rm pc} \times 1 {\rm pc}$) at C-band. The second, northeastern, source (hereafter NE) is more extended and displaced by $\sim$30\,pc from the first one at P.A.$\sim$70\degr.

Table~\ref{table:evn_cont} displays, for each source and frequency, coordinates, dimensions, peak and integrated flux densities. All parameters were determined fitting a 2-dimensional Gaussian with AIPS task JMFIT, except for the NE source that is clearly resolved at 5\,GHz and could not be approximated by a simple Gaussian. In this case, we derived the peak position and flux density using AIPS task IMSTAT. The integrated flux density, instead, was obtained using the task BLSUM over a polygon enclosing the emission down to the 5$\sigma$ level. We estimated the size of the source directly from the map, measuring the extension of the radio emission above the 5$\sigma$ level, in two orthogonal directions. The task JMFIT directly provides the errors on the peak and integrated flux density. For the NE source at 5\,GHz, we adopted the rms noise of the map as the uncertainty on the peak flux density, while the error on the integrated flux density was calculated using the formula reported in \citet[][their Sect.~3]{panessa2015}. The uncertainty in the absolute positions is dominated by the error on the position of the phase calibrator J1555-0326, 0.2 and 0.4\,mas in right ascension and declination, respectively. Table~\ref{table:evn_cont} also includes estimates of the brightness temperatures and radio powers obtained from the integrated flux densities and sizes of the sources.

We found that the positions of the sources at the two frequencies are not fully consistent. Indeed, the 5\,GHz positions are offset by $\sim$4\,mas for the most compact component and $\sim$7\,mas for the more extended one, in the same direction (i.\,e., toward the northeast). As discussed in Appendix~\ref{app:shift}, we believe that this misalignment is more likely due to a phase error rather than to a real astrophysical phenomenon. Therefore, for the purpose of comparing the maps, we realigned them shifting the L-band image by an amount necessary to make the centroid of component SW coincident at the two frequencies.  

\begin{figure*}
\includegraphics[scale=0.45]{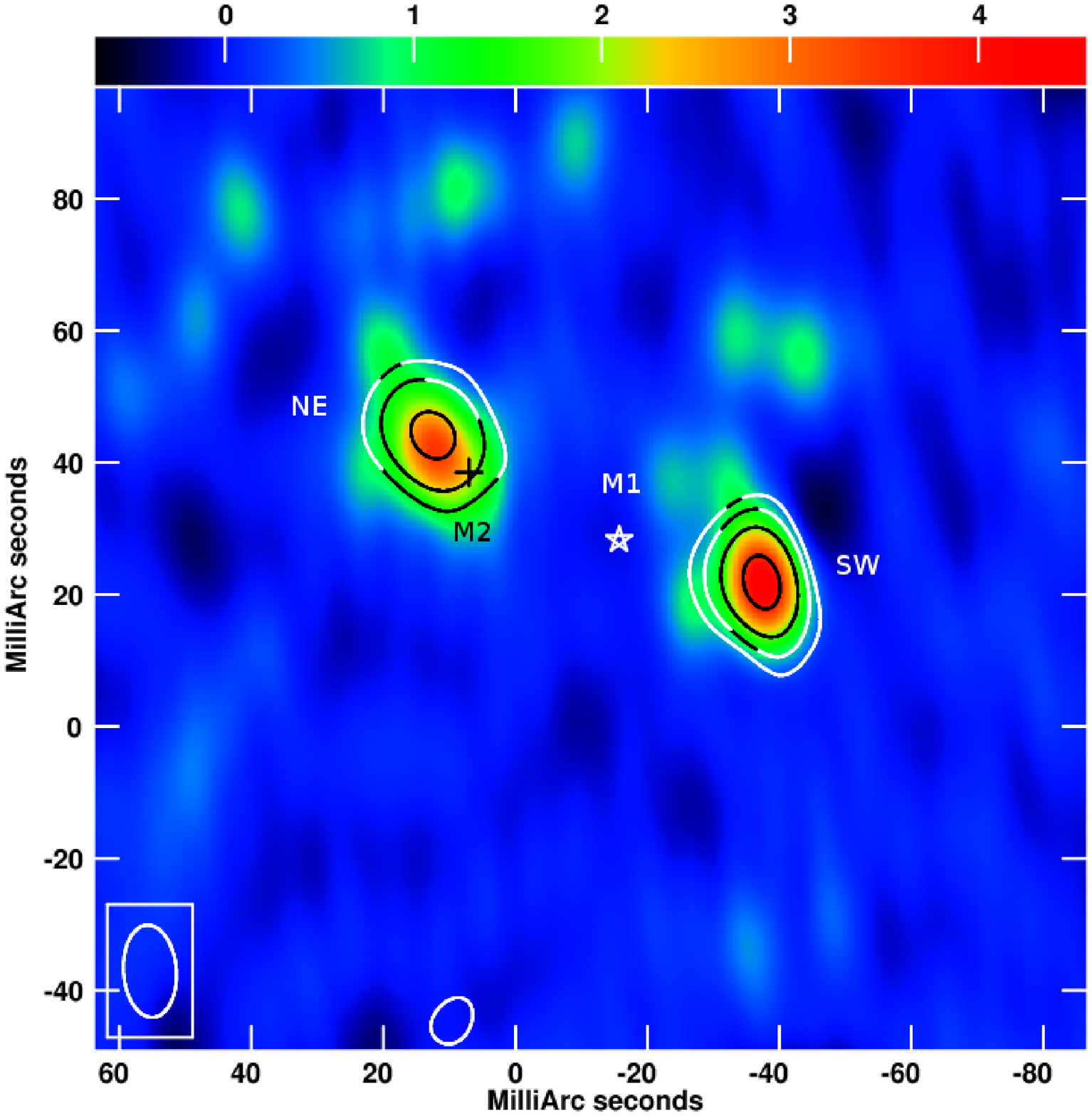}
\includegraphics[scale=0.45]{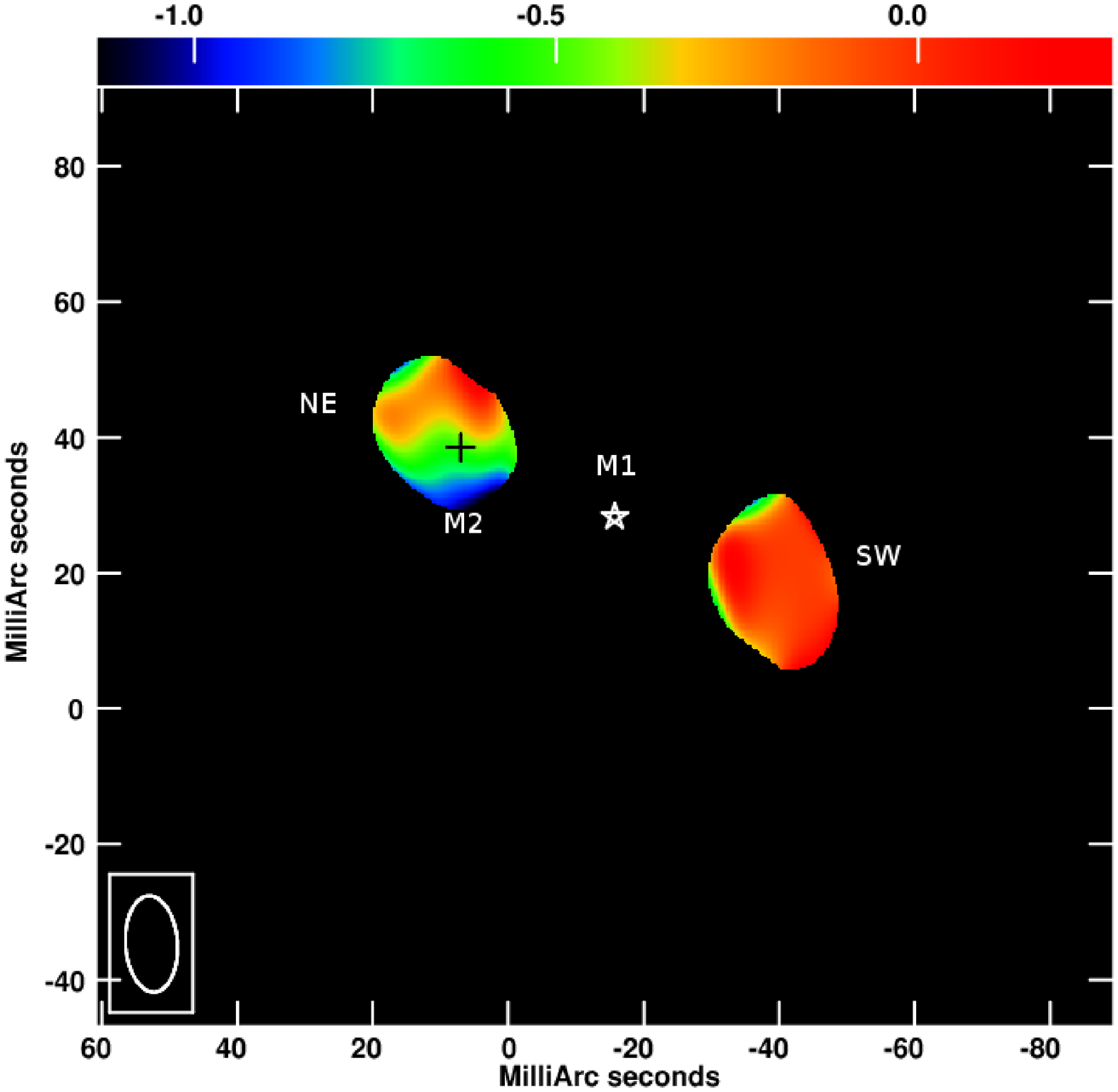}
\caption{Radio continuum emission in the nucleus of IRAS15480 ({\it left}). The color scale represents emission at L-band (shifted as described in Appendix~\ref{app:shift}), ranging from -0.7 to 4.6\,mJy/beam, while the overlaid contours delineate the C-band emission convolved with the L-band beam (contour levels are -1, 1, 2, 4, 8, 16, 32, 64 $\times$ 0.45 mJy/beam). Spectral index map of the nuclear region in IRAS15480 ({\it right}). The map shows only the emission regions above the 5$\sigma$ noise level. The positions of the water maser spots detected with the VLBA are also indicated. The star and the cross mark the location of the narrow (M1) and the broad blueshifted line emission (M2), respectively \citep[for details see][]{castangia2016}.} 
\label{fig:spix}
\end{figure*}

In order to compute spectral indices, we also convolved the C-band map with the L-band beam. We then derived the spectral indices between 1.7 and 5\,GHz following the convention $S \propto \nu^{\alpha}$. For the most compact component, SW, we obtained a rather flat spectral index, $\alpha_{5, p}^{1.7} =0.01 \pm 0.02$ and $\alpha_{5, i}^{1.7} =-0.06 \pm 0.04$, using peak and integrated flux densities, respectively. Source NE, instead, is characterized by a steeper spectrum with $\alpha_{5, p}^{1.7} =-0.39 \pm 0.04$ and $\alpha_{5, i}^{1.7}=-0.56 \pm 0.06$. Since the latter component is resolved, we have also produced a spectral index map, which is shown in Fig.~\ref{fig:spix}.

Neither of the continuum sources we detected in our EVN maps were seen in previous VLBA images of IRAS15480 at C- and K-band \citep[see][for details]{castangia2016}. The non-detection at C-band can be explained by the low sensitivity of the VLBA observation (5$\sigma$ rms was 3.5\,mJy\,beam$^{-1}$). The reason why we do not see any of the sources in the 22\,GHz image is not straightforward. Considering the flux densities in Table~\ref{table:evn_cont} and the 1.7-5\,GHz spectral indices, we expect 22\,GHz integrated flux densities of 1.15\,mJy and 5.1\,mJy for sources NE and SW, respectively, that should have been detected in our VLBA maps. However, given the smaller beam of the K-band continuum image ($\sim 2 {\rm mas} \times 1 {\rm mas}$), both sources might be resolved at K-band. If source NE was resolved in two or three components, it could easily fall below the 5$\sigma$ detection threshold of 0.5\,mJy of the K-band continuum image. The same does not hold for source SW. Indeed, if we consider the size of SW, 12\,mas$^2$, and the area subtended by the K-band beam ($\sim$2\,mas$^2$), this source can be resolved into, at maximum, 6 components, each with a flux density of 0.85\,mJy ($\sim 8 \sigma$ in the K-band map). Therefore, we should have seen source SW in the VLBA K-band continuum image. If we,  conservatively, use the peak flux densities, instead of the integrated ones, we obtain 22\,GHz peak flux densities of 0.35\,mJy\,beam$^{-1}$ and 2.3\,mJy\,beam$^{-1}$, for sources NE and SW, respectively. Hence, we should have detected SW. The fact that it was not detected suggests that the spectrum of SW becomes steeper at frequencies higher than 5\,GHz. 
 

\begin{table*}
\caption{Parameters of the nuclear continuum sources in IRAS15480: coordinates, deconvolved sizes, position angles, peak and integrated flux densities, brightness temperatures, and radio powers.}      
\label{table:evn_cont}      
\centering          
\begin{tabular}{cccccccccc}     % 9 columns 
\hline\hline       
Label & $\nu$ & RA                     & Dec.               &  Deconvolved Size & P.A.    & $S_{\rm peak}$        & $S_{\rm int}$ & $T_{\rm B}$  &  $P$ \\
      &       &  \multicolumn{2}{c}{J2000}                 &                   &         &                     &             &            & \\
      & (GHz) & 15$^{\rm h}$ 50$^{\rm m}$ & -03\degr 53\arcmin & (mas), [pc]       & (\degr) & (mJy\,beam$^{-1}$)   & (mJy)       & (K)        & (erg\,s$^{-1}$\,Hz$^{-1}$) \\
\hline
NE    & 1.7        & 41$^{\rm s}$.49860  & 18\arcsec.0115     & 15$\times$7, [9$\times$4] & 52 & 3.2$\pm$0.1   & 7.1$\pm$0.3 & 3.9$\times$10$^7$ & 1.4$\times$10$^{29}$ \\
      & 5.0        & 41$^{\rm s}$.49892  & 18\arcsec.0068     & 15$\times$10, [9$\times$6] & -- & 0.64$\pm$0.04 & 2.8$\pm$0.1 & 6.0$\times$10$^5$ & 5.6$\times$10$^{28}$ \\ 
SW    & 1.7        & 41$^{\rm s}$.49532  & 18\arcsec.0316     & 7$\times$4, [4$\times$2] & 50 & 4.5$\pm$0.1   & 6.0$\pm$0.2 & 1.6$\times$10$^8$ & 1.2$\times$10$^{29}$ \\
      & 5.0        & 41$^{\rm s}$.49546  & 18\arcsec.0284     & 6$\times$2, [4$\times$1] & 69 & 2.30$\pm$0.03 & 5.6$\pm$0.1 & 2.3$\times$10$^7$ & 1.1$\times$10$^{29}$ \\
\hline                  
\end{tabular}
\end{table*}

\section{Discussion}\label{sect:discussion}

\subsection{The high detection rate}\label{sect:ctsample_discu}
With the detection of IRAS15480, the maser detection rate of the entire sample reaches 50\%, which is one of the highest ever found in maser surveys. \citet{henkel05} searched for 22\,GHz H$_2$O masers in two classes of objects: a small (14 objects) sample of Seyferts with known jet-Narrow Line Region interactions and a larger (45 objects) sample of FIR bright galaxies. The former sample yielded a detection rate comparable to the one we found \citep[7/14, i.\,e. 50\%][]{henkel05}. In general, however, maser detection rates are typically a few percent \citep{braatz2018}. In order to investigate if such a high detection rate might be due to a distance bias, we plot in Fig~\ref{fig:detrate} the redshift distribution of the galaxies in our sample (upper panel) and the maser fraction in the same redshift bins (lower panel). The large majority of the objects (90\%) are in the first 4 redshift bins (0.002$\leq z <$0.04). In the case of a distance bias, one would expect the maser fraction to decrease with increasing redshift, while Fig~\ref{fig:detrate} shows that the detection rate is consistent with 50\% in the whole redshift range. Furthermore, given the small number of galaxies with z$>$0.04 (only 3), the absence of maser detections at large redshift is more likely due to the paucity of observable targets than to a distance bias. Finally, we notice that the redshift distribution of our sample extends to much higher values compared to other samples of AGN searched for 22\,GHz maser emission. For example, the galaxies in the FIR-sample of \citet{henkel05} that has a comparable number of objects, have z$<$0.02 and yield a detection rate of 22\%, lower than the value we obtain in the same redshift interval (13/22, i.\,e. $\sim$60\%). In the complete samples of Seyfert galaxies studied by \citet{panessa2013}, the largest redshift is 0.0034 and the detection rate 26\%. Although a comparison with another sample with the same redshift distribution would be necessary to verify our conclusion, we believe that the large maser fraction found in our CT AGN sample is a result of an efficient selection method that favours the detection of maser emission and that the distance bias, if present, is negligible. 

The average luminosity of the known water masers in the CT AGN sample (Table~\ref{table:sample}), is $\sim$190\,L$_\sun$. Using Eq.~\ref{eq:lum} and assuming a typical 1$\sigma$ rms of 10\,mJy per channel (Table~\ref{table:sample}, where we did not considered the unusually large rms of AM1925-724) we calculate that a maser with a typical luminosity of 190\,L$_\odot$ would be detectable up to a distance of $\sim$130\,Mpc ($z \la 0.03$) if the full width at half maximum (FWHM) of the line is 10\,km\,s$^{-1}$ and up to 40\,Mpc ($z \la 0.009$) if the same luminosity is emitted in a broad line, with FWHM=100\,km\,s$^{-1}$. Very broad lines and narrow features at redshifts larger than 0.03 may be undetected because of the limited sensitivity of the surveys. Therefore, the overall detection rate might be even higher than 50\%, possibly reaching the value obtained for the first redshift bins. A detailed statistical analysis of the CT AGN sample, including this issue, will be the subject of a forthcoming paper (PaperIII).
{{\begin{figure}
\includegraphics[scale=0.4]{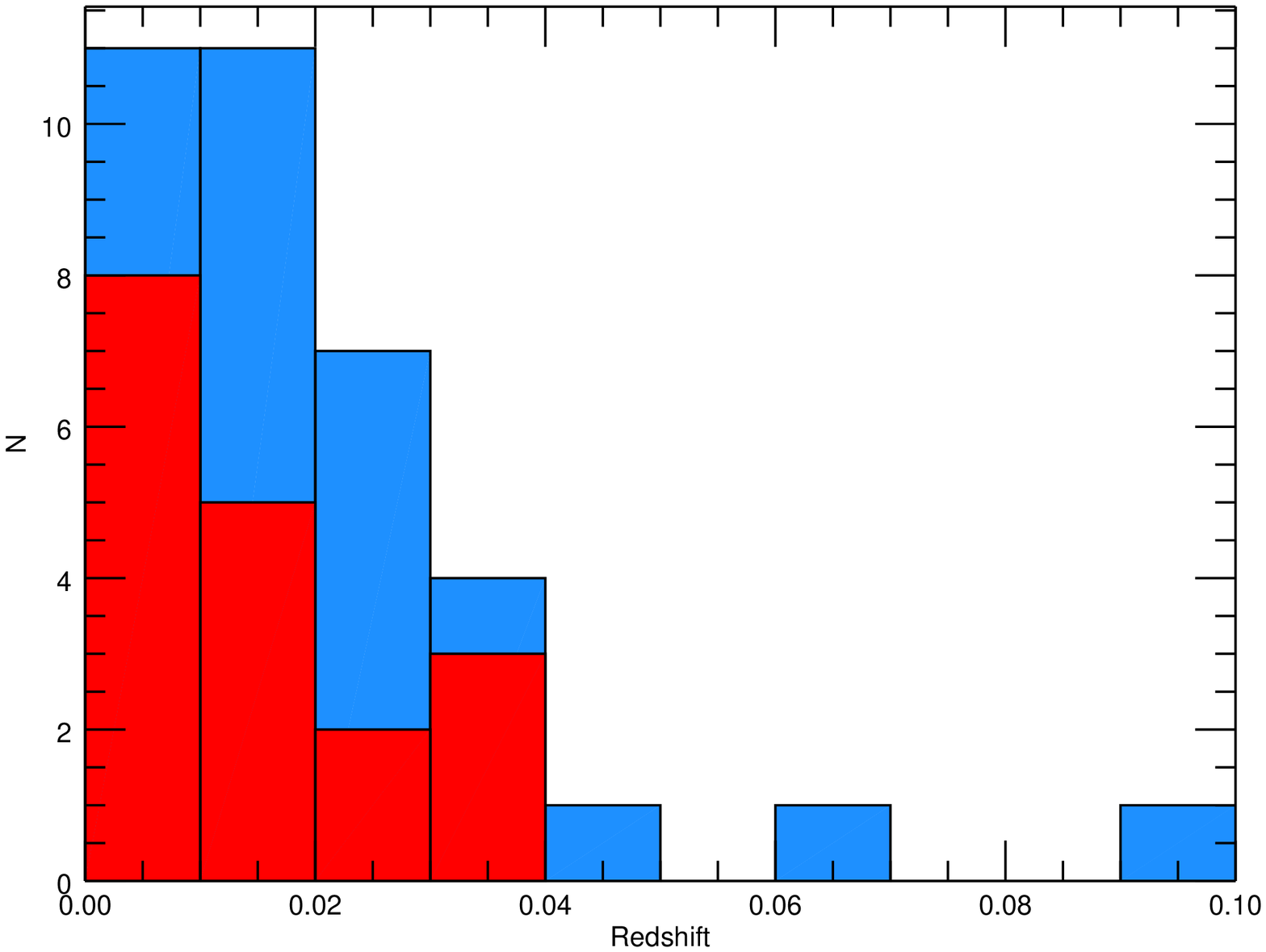}
\includegraphics[scale=0.4]{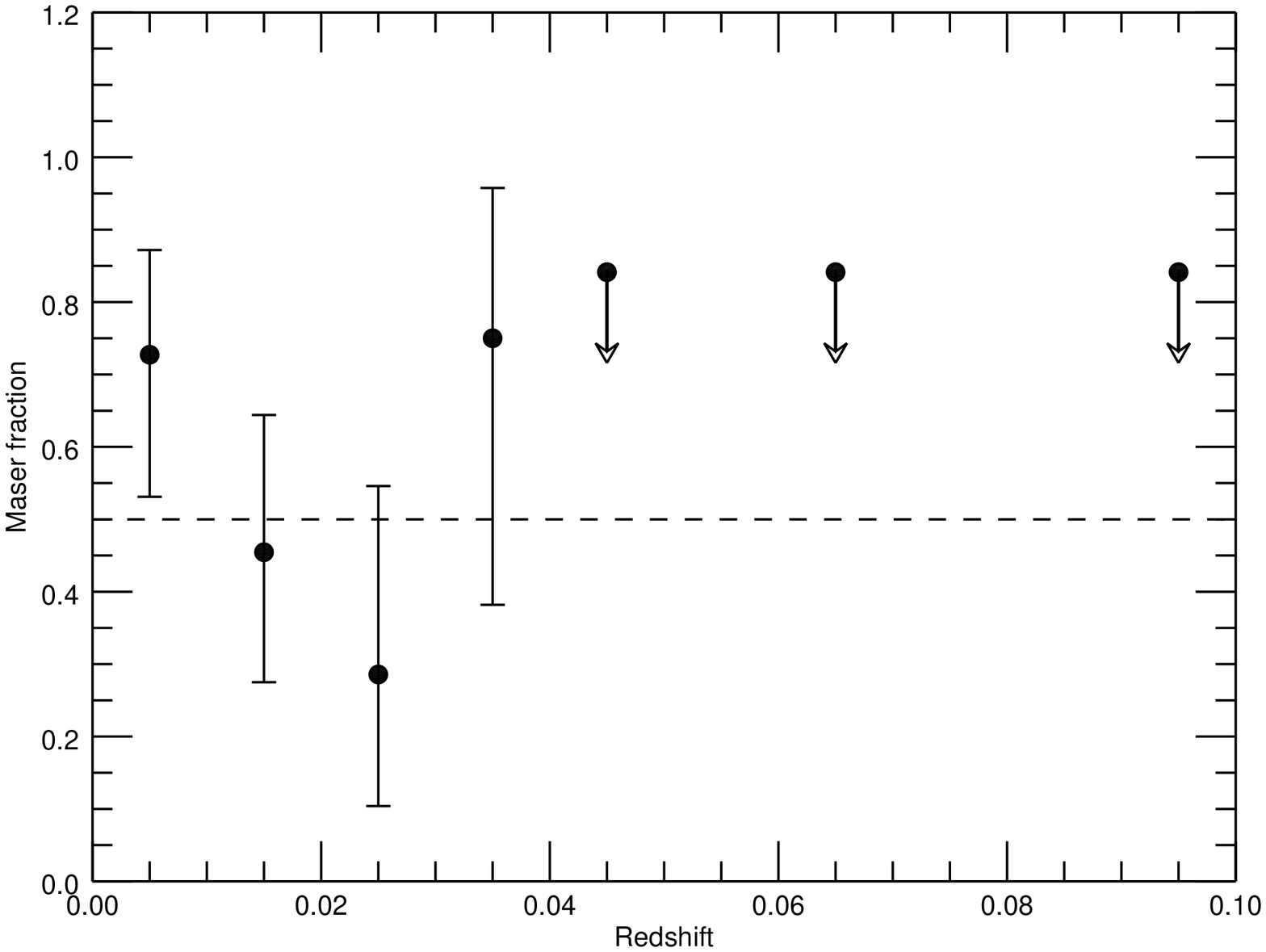}
\caption{{\it Upper panel:} redshift distribution of the galaxies in the CT AGN sample (blue) and of the known water masers in the same sample (red). {\it Bottom panel:} Water maser fraction in the same redshift bins of the upper panel. The dashed line indicates the maser fraction of the whole sample. Errors and upper limits are based on the binomial distribution and are calculated following \citet{gehrels86}.}
\label{fig:detrate}
\end{figure}

%luminosità media (escluse M51 e UGC5101)
%110 Lsun
%considerando una larghezza che vada da 10 a 100 km/s e un rms medio di 10 mJy (escludendo AM1925-724)
%otteniamo
%D = 98 Mpc --> z = 0.023 (FWHM = 10 km/s)
%D = 31 Mpc --> z = 0.007 (FWHM = 100 km/s)
%per il maser più debole (M51, con 2 Lsun e FWHM = 10 km/s, Ho et al. 1987), otteniamo che 
%con un rms per canale di 1 km/s, questo maser si potrebbe vedere sino ad una distanza di:
%D = 13 Mpc (z=0.003)
%quello più forte invece (UGC5101, con 1762 Lsun e FWHM = 270 km/s, Zhang et al. 2006)
%si potrebbe vedere sino ad una distanza di:
%D = 75 Mpc (z=0.017)

\subsection{The nuclear radio continuum emission in IRAS15480}\label{sect:cont_discu}
The total flux density of IRAS15480 measured with the EVN amounts to 13.1$\pm$0.4\,mJy and 8.4$\pm$0.1\,mJy at 1.7 and 5\,GHz, respectively. Thus, at L-band, we recovered only 30\% of the VLA flux density \citep[$S_\mathrm{1.4}$=42\,mJy, NVSS;][]{condon98}. This implies that most of the radio emission at this frequency is not concentrated in a compact nuclear source but is spread over a region larger than 0\farcs1 (60\,pc), the EVN largest detectable angular scale (corresponding to the shortest baseline, Ef-Wb). At the angular resolution of the NVSS survey (full width at half maximum beam-size $\sim$45\arcsec), the radio emission appears to be unresolved and is, therefore, confined within a sphere of $\sim$10\,kpc diameter. 

The fact that the parsec-scale radio emission accounts for only a small fraction of the flux density measured in lower resolution images, is a common feature in Seyfert galaxies \citep[e.\,g.,][]{orienti2010,panessa2013}. However, the origin of the missing flux is difficult to assess. Most of the kpc-scale radio emission from normal galaxies (i.\,e., galaxies that do not harbour an AGN) is synchrotron radiation from relativistic electrons produced by supernova renmants (SNR) and free-free emission from H{\sc ii} regions \citep{condon92}, with the former dominating at low frequency. However, in galaxies hosting an active nucleus, the source of this diffuse emission might also be the AGN itself, through thermal emission from AGN ionized gas or synchrotron emission from a disrupted jet \citep[][and references therein]{orienti2010}. Understanding the origin of the missing flux is beyond the scope of this work. Furthermore, a period of about 20 years has passed between the NVSS and our EVN observations. This long time-lag makes the comparison of the VLA and VLBI flux densities controversial, because significant variations (with time-scales of years) in the nuclear radio flux density of Seyferts and other radio quiet AGN have been often found \citep[e.\,g.,][]{mundell09,doi2013}. 

No hints of large or kpc scale jet-like structures are visible at the intermediate resolution of the VLA X-band map that shows an unresolved nuclear radio source with an angular diameter $<$0.1\arcsec, corresponding to $<$60\,pc \citep{thean00,schmitt01}. However, although the potential jet in IRAS15480 is confined to scales of tens of parsecs, the monochromatic radio luminosity of IRAS15480 measured on VLBI scales exceeds 10$^{29}\,$erg\,s$^{-1}$\,Hz$^{-1}$ at both 1.7 and at 5\,GHz, placing it among the most powerful Seyfert galaxies in the radio band (e.\,g., NGC\,3079, NGC\,4278, IC\,5063). Interestingly, the radio luminosity of IRAS15480 is consistent with the critical threshold introduced by \citet[][see their Fig.~3]{liu2017} to detect water maser emission. Indeed, confirming an earlier conclusion by \citet{zhang2012}, \citet{liu2017} found that maser galaxies tend to have higher radio luminosities by a factor of 2--3 than non-masing ones. 
%Aggiungere confronto con altre Seyfert luminose (N1068,N3079,N4151,N4278,N4579,IC5063).
%N1068
%N4151 (Ulvestad et al. 2005 and references therein)
%Jet extent --> 300 pc
%Jet power --> 8x10^20 W Hz^-1 o 4x10^37 erg s^-1 a 5 GHz
%The nucleus has a flat spectrum, Tb=2.1x10^8 K, D<0.1 pc and P=2x10^20 W Hz^-1
%Changes in position angle from the inner parsecs to the 100 pc scale
%Non relativistic jet
%N4278 (Giroletti et al. 2005)
%Jet extent --> 1.4 pc
%Jet power --> 3x10^21 W Hz^-1 o 1.4x10^38 erg s^-1 a 5 GHz
%S-shaped jets emerging from a flat-spectrum core 
%Mildly relativistic jet
Using the EVN flux densities we derive a 5\,GHz nuclear radio continuum luminosity of 8.6$\times 10^{38}$\,erg\,s$^{-1}$. Dividing the radio luminosity by the unabsorbed X-ray luminosity in the 2--10\,keV band \citep[$L_\mathrm{2-10keV}=1-4\times10^{43}$\,erg\,s$^{-1}$;][]{brightman2011}, we then obtain a value of $\log R_\mathrm{X}$ between $-$4.1 and $-$4.7. These values bracket the approximate boundary established by \citet{terashima03} between radio loud and radio quiet AGN ($\log R_\mathrm{X}=-4.5$).

In the following we discuss the nature of the radio continuum components found in the nuclear region of IRAS15480 in the framework of the radio emission in radio quiet AGN. A star formation origin is unlikely, due to the relatively large radio power yielding very high brightness temperatures (Table~\ref{table:evn_cont}) and will therefore not be discussed below. 
%Aggiungere stima del rapporto standard L5/LB? 
%Aggiungere paragrafo su emissione radio nelle galassie radio quiet

\subsubsection{The SW component}\label{sect:cont_sw}
The most compact component, SW, is characterized by a flat spectral index (Sect.~\ref{sect:results}) and brightness temperatures of 1.6$\times$10$^8$\,K and 2.3$\times$10$^7$\,K, at 1.7\,GHz and 5\,GHz, respectively {\bf (Table~\ref{table:evn_cont})}. Flat-spectrum nuclear radio sources might be produced by synchrotron self absorbed emission from the base of a jet; this is the typical case in radio loud AGN and in an increasing number of radio quiet objects \citep[e.\,g.,][]{mundell00,caccianiga01,giroletti09,bontempi2012}. Alternatively, it might also be interpreted as thermal bremsstrahlung emission from an X-ray heated corona or wind arising from the accretion disc, as was proposed for the flat-spectrum source S1 in the nucleus of NGC\,1068 \citep{gallimore97} and, possibly, NGC\,4388 and NGC\,4477 \citep{mundell00,bontempi2012}. However, the large values of $T_\mathrm{B}$ rule out a thermal mechanism as the origin of the emission for component SW. For comparison, the brightness temperature of source S1 in NGC\,1068 is in the range 10$^5$--\,4$\times$10$^6$\,K \citep{gallimore97}. On the other hand, the properties of SW ($T_{\rm b}>10^8$\,K and $\alpha\sim 0$) are consistent with synchrotron self absorbed emission from a radio core. The size of SW, which extends for about 4\,pc along P.A.$\sim$70{\degr} (Table~\ref{table:evn_cont}), is, however, larger then those of typical radio cores, usually unresolved with sizes of less than 1\,pc \citep[e.\,g.,][]{ulvestad05}. In addition, although the significance is low (5$\sigma$), a weaker extension towards east is visible at both 1.7\,GHz and 5\,GHz (Fig.~\ref{fig:cont+maser}). While the elongation along P.A.$\sim$70{\degr} is in the direction of component NE and, therefore, is consistent with SW representing the initial part of a jet, the weaker extension along P.A.$\sim$90{\degr} is more difficult to explain within this scenario. 

If we assume that the flat spectrum of component SW is due to synchrotron self absorption, we can calculate the magnetic field $B$ necessary to have the turnover frequency between 1.7 and 5\,GHz, using the formula derived from \citet{kellermann81}, for low redshift objects:
\begin{equation}
B[{\rm G}]\sim\frac{\nu_{\rm p}^5[{\rm GHz^5}]\theta^4[{\rm mas^4}]}{f(\gamma)^5 S_{\rm p}^2[{\rm Jy^2}]},  
\end{equation}
where $\nu_{\rm p}$ is the peak or turnover frequency in GHz, $\theta$ is the source diameter in mas, and $S_{\rm p}$ is the flux density at the turnover frequency. $f(\gamma)$ is a function that weakly depends on the spectral index $\gamma$ of the energy distribution of electrons and is about 8 for a typical value $\gamma\sim$2. Using sizes and fluxes from Table~\ref{table:evn_cont}, we obtain that a turnover frequency between 1.7 and 5\,GHz would imply a magnetic field in the range $10^3-10^6$\,G. Since magnetic fields in compact radio sources typically are of the order of a few mG \citep[e.\,g.,][]{kellermann81}, the very high value found by us indicates that syncrothron self absorption cannot be the mechanism that flattens the spectrum of component SW. On the other hand, a spectral index $\alpha\sim 0$ might be caused also by free-free absorption of an intrinsic synchrotron source. We propose that, similarly to component NE (see next subsection), SW may represent a jet knot, where synchrotron emission is absorbed by the thermal ionized gas produced at the shock front where the jet impacts the ambient medium. 
 
%Può trattarsi di un outflow invece (o in aggiunta) a un jet??? 
%La differenza dovrebbe essere che l'outflow non è collimato, ma ha anche una Tb minore???--> Non lo so e non trovo niente in letteratura 
%Può trattarsi di assorbimento free-free anziché autoassorbimento? Si può stabilire se gli elettroni sono o no relativistici??
%Sì, può trattarsi di assorbimento free free (vedi modello di Bicknell et al. (1997) applicato alla componente C in N1068)
%Nel paper di Mundell et al. (2000) c'è la formula Tb=2x10^9gamma K, dove gamma è il fattore di Lorentz, da qui si vede che per gamma=1 (v<<c, particelle non relativistiche), Tb è già 2x10^9K, 
%quindi fino a questi valori di Tb non è necessario che gli elettroni siano relativistici. Questo sarebbe anche il nostro caso!
%Aggiungere stima della densità di energia e del campo magnetico di equipartizione come in Bontempi et al. 2012? 
%Fatto! I valori sono in rs_physics.dat
%Umin=1.7 e 3.6 x 10^5 erg/cm^3, rispettivamente in banda L e C 
%Beq=1.4 e 2.0 mG, rispettivamente in banda L e C 

\subsubsection{The NE component}\label{sect:cont_ne}
Differently from source SW, the more extended component NE has a steep spectrum between 1.7 and 5\,GHz (Sect.~\ref{sect:results}), compatible with optically thin synchrotron emission, and a lower brightness temperature ($T_\mathrm{B}\sim 4\times$10$^7$\,K and $\sim 5\times$10$^5$ K, at 1.7 and 5\,GHz, respectively). It is clearly resolved at both frequencies, with dimensions of $9 {\rm pc} \times 4  {\rm pc}$. The characteristics of the component NE (i.\,e., steep spectrum, parsec-size, and radio power in excess of 10$^{28}$\,erg\,s$^{-1}$\,Hz$^{-1}$), are consistent with those of a weak jet-knot or lobe. Interestingly, the spectral index map in Fig.~\ref{fig:spix} (right panel) reveals a gradient in the $\alpha$ distribution. Indeed, a region with a flat spectral index ($\alpha\sim0$) is present in the northern part, then the spectrum steepens toward the south reaching $\alpha<-1$. This distribution suggests a jet-shock scenario, as the one proposed, for example, for components C and S2 in the nuclear region of NGC\,1068 \citep{gallimore04}. \citet{bicknell97} demonstrated that fast radiative shocks produced by the interaction of the jets with a dense interstellar medium are able to ionize the gas, creating an ionized envelope with very high emission measures. This gives rise to free-free absorption of the syncrothron emission from the jet and, hence, to a local flattening of the spectrum at the shock front. Although this model was developed by \citet{bicknell97} to explain the low-frequency turnover in the radio spectra of compact radio loud objects (such as gigahertz peaked spectrum {\bf (GPS)} sources), it is also able to account for the properties of the compact radio sources in the nucleus of the radio quiet Sy~2 NGC\,1068 \citep{gallimore04}. The detection of H$_2$O maser emission, typically collisionally pumped \citep[e.\,g.][and references therein]{lo05}, at the location of component NE (Fig.~\ref{fig:spix}), is an additional evidence for the presence of a shock.
%Aggiungere stima della densità di energia e del campo magnetico di equipartizione come in Bontempi et al. 2012? 
%Fatto! I valori sono in rs_physics.dat
%Umin=5.4 e 2.0 x 10^4 erg/cm^3, rispettivamente in banda L e C 
%Beq=0.8 e 0.5 mG, rispettivamente in banda L e C

\subsection{Origin of the water maser in IRAS15480}\label{sect:maser_discu}

In \citet{castangia2016} we analysed 22\,GHz single-dish spectra taken with the GBT at three epochs and interferometric VLBA observations of the luminous water maser in IRAS15480.
On the basis of these data, we suggested a composite nature for the maser emission. The broad component (M2), likely originates from the interaction of a radio jet with ambient molecular clouds. We favoured an outflow origin for the narrow feature (M1). From our new EVN radio continuum images we can associate the location of the line emission with sources of nuclear activity, revealing the presence of a compact radio jet in IRAS15480. We favour the interpretation of the two compact continuum sources, SW and NE, as jet knots. In this scenario, the position of the putative nucleus should be along the imaginary line connecting the two radio continuum sources. The narrow line spot M1, being nearly equidistant from SW and NE, likely coincides with the position of the core and, hence, may be associated with the accretion disc or a nuclear outflow. The broad line emission, whose position coincides with source NE, is likely associated with a jet-cloud interaction as in NGC\,1068 \citep{gallimore01} and Mrk348 \citep{peck03}.

%\begin{enumerate}
%\item SW and NE are both jet knots.
%In this scenario, the position of the putative nucleus should be along the imaginary line connecting the two radio continuum sources.
%The narrow line spot M1, being nearly equidistant from SW and NE, likely coincides with the position of the core and, hence, may be associated with the accretion disc or a nuclear outflow.
%The broad line emission, whose position coincides with source NE, is likely associated with a jet-cloud interaction as in NGC\,1068 and Mrk348 \citep{gallimore01,peck03}.
%\item SW is the jet-base and NE a jet knot.  
%In this picture, the position of the putative accretion disc/torus should coincide with source SW.
%Both maser spots are along the direction of propagation of the jet at distances of $\sim$15 and 30\,pc from the core/accretion disc position, therefore, they can be both associated with the jet, the spot corresponding to the broad line with a jet-cloud interaction and M1 with a molecular outflow produced by the jet \citep{morganti2015}.
%\end{enumerate}
%Dettagli sui due scenari
%Se M1 è un disc-maser perché non vediamo le righe satelliti?
%Spiegare meglio il caso dell'outflow generato dal jet
If M1 is associated with an edge-on accretion disc, we should see two additional groups of lines displaced by several km\,s$^{-1}$ from the systemic velocity and bracketing it, as typically seen in disc-masers \citep[e.\,g.,][]{miyoshi95,pesce2015}. However, these are not observed. The satellite lines could be below the detection threshold of the single-dish spectrum \citep[e.\,g. NGC~2639;][]{wilson95} or, instead, be intrinsically absent as is expected in the case of inclined water maser discs \citep{darling2017}.
The position of M2, coincident with the optically thin region of source NE (Fig.~\ref{fig:spix}, right panel), instead, is in agreement with the picture proposed by \citet{peck03} to explain the maser emission in Mrk348. According to these authors, the masing region should be located within or immediately behind the radiative shock which is thought to precede the expansion of the jet into the interstellar medium (see their Fig.~11). If, as supposed in Sect.~\ref{sect:cont_ne}, the flat spectrum area of component NE marks the ionized zone at the shock front, then the maser spot M2 likely arises from the postshock region. Differently from \citet{peck03}, our data do not allow us to constrain the jet properties (density and velocity). However, we can infer an upper and lower limit for the shock velocity.
Assuming that the jet is oriented close to the plane of the sky, the maser emission is more likely to arise from the head of the expanding cocoon driven into the ISM by the jet. Indeed, regardless of the type of shock, $C$ shocks or fast dissociative $J$ shocks \citep[for a definition of the shock types see][]{draine80}, maser radiation is preferentially beamed perpendicular to the motion of the emitting gas (i.\,e., perpendicular to the shock velocity), so that brighter masers will have line of sight (or radial) velocities lower than their transverse velocities \citep[e.\,g.,][]{hollenbach2013}.Within this scenario, we assume that the component of the shock velocity in the direction of the jet propagation is equal or larger than the velocity separation, along the line of sight, between the masing material and the ambient (preshock) gas. Therefore, a lower limit to the shock velocity is set by the velocity offset of the maser emission with respect to the systemic velocity, $\sim$120\,km\,s$^{-1}$ w.r.t $V_{\rm sys}$ for M2. In order for H$_2$O to reform in the postshock gas the dust grains must survive the shock passage. Hence, the shock velocity must be $\leq$300\,km\,s$^{-1}$. For higher velocities the dust grains are destroyed and it is difficult for the water molecules to reform directly in the gas phase \citep{elitzur89}. Obviously, in case the maser emission does not come from the head of the expanding bubble but is produced laterally, where the speed is lower, the shock velocity in the direction of the jet propagation can be also higher than 300\,km\,s$^{-1}$. The same holds also in case the jet is inclined with respect to the plane of the sky.  

%N1068: blueshifted (-130 e -195) at 30pc from the core, only VLA observations at 100 mas resolution (7 pc)  
%Mrk348: redshifted (+135 km/s) at <1pc from the core, one unresolved spot (<0.25 pc) FWHM=140 km/s  
%Spiegare perché non è un caso simile a quello di N1052 e perché vediamo solo un a riga sottile invece di una larga...

\section{Conclusions}\label{sect:conclusions}

We searched for 22\,GHz water maser emission in a well defined sample of 36 CT AGN, selected in the local Universe through a combination of mid-IR (\textit{IRAS}) and X-ray (\textit{XMM-Newton}) data. When we compare the CT AGN sample with other H$_2$O surveys, it shows one of the highest detection rates ever found. Indeed, including previously detected sources, with the newly discovered megamaser ($L_{\rm H2O} \sim 200$\,L$_\odot$) in the Seyfert~2 galaxy IRAS15480, the maser detection rate of the CT AGN sample is 50\% (18/36). We also observed the radio continuum emission in IRAS15480 at parsec-scale resolution, using the European VLBI Network. These high angular resolution observations were performed in order to associate the position of the luminous water maser with sources of activity in the nuclear region, thus shedding light on the origin of the line emission. Our interferometric data reveal that: 

\begin{itemize}
\item IRAS15480 is one of the most powerful Seyfert galaxies (which are usually radio quiet) in the radio band, with a monochromatic luminosity exceeding 10$^{29}\,$erg\,s$^{-1}$\,Hz$^{-1}$ at VLBI scales. This is consistent with the critical threshold found by \citet{liu2017} to detect water maser emission.
%At L-band (1.7\,GHz) we recovered only 30\% of the flux density measured in lower resolution (NVSS) images. This suggests that most of the radio emission at this frequency is not concentrated in a compact nuclear source but is spread over a region larger than $\sim$60\,pc. We estimate that a substantial contribution (at least 50\%) to the missing flux is due to the AGN activity. We suggest that, as in the case of other Seyferts (e.\,g., NGC4151), a disrupted jet might produce a steep-spectrum low-surface  brightness emission that cannot be detected with VLBI observations because of sensitivity/technical limitations.      
\item The radio continuum emission in the nucleus of IRAS15480 is resolved into two bright components: a compact ($4 {\rm pc} \times 1 {\rm pc}$), slightly resolved, source in the south-west (SW) and a more extended ($9 {\rm pc} \times 4 {\rm pc}$) component in the north-east (NE), displaced by $\sim$30\,pc from the first one. The properties of these sources (spectral indices, brightness temperatures, dimensions, and radio power) indicate that their radio emission is synchrotron radiation, most likely produced by two weak jet knots of a compact radio jet. Both components show evidences {\bf for} strong interaction with a dense interstellar medium.
\item The position of the narrow maser line, M1, is along the imaginary line connecting the two radio continuum sources, SW e NE, and is nearly equidistant from them. Hence, it might trace the position of the core (not visible in the radio continuum images) and be associated with the accretion disc or a nuclear outflow. The position of the broad maser feature, M2, instead, coincides with the optically thin part of source NE, suggesting that the maser emission might originate within or immediately behind the shock produced by the interaction of the jet with the interstellar medium, as it was proposed for the maser in Mrk~348. Within this scenario, we infer a lower limit to the shock velocity of $\sim$120\,km\,s$^{-1}$, consistent with fast $J$ shocks.  
\end{itemize}

The EVN maps, therefore, confirm our initial hypothesis that part of the maser emission is produced by the impact of a radio jet with molecular clouds in the host galaxy, adding a new source to the few confirmed jet-masers reported so far. In addition, the combination of VLBI radio continuum and maser observations unveil the presence of a compact radio jet and of strong interactions of the latter with the dense interstellar medium in the nucleus of a relatively radio quiet galaxy. This highlights the potential of maser studies to shed light on the parsec scale environment around AGN and, possibly, on the role of low power jets on galaxy evolution.

\begin{acknowledgements}
We wish to thank the Sardinia Radio Telescope (SRT) Operations Team and, in particular, C. Migoni, A. Melis, and F. Gaudiomonte, for their help with the L-band session at the SRT. We are also grateful to the anonymous referee for his/her useful suggestions. The European VLBI Network is a joint facility of independent European, African, Asian, and North American radio astronomy institutes. Scientific results from data presented in this publication are derived from the following EVN project code: EC047. 
\end{acknowledgements}

%-------------------------------------------------------------------

\bibliographystyle{aa} % style aa.bst
\bibliography{castangia2019} % your references Yourfile.bib

\appendix
\section{Misalignment between L- and C-band images}\label{app:shift}

Fig.~\ref{fig:shift} displays the radio continuum emission at 5\,GHz overlaid onto the 1.7\,GHz emission. The overlapping highlights a possible misalignment between the two maps. Indeed, in the C-band image, both continuum components appear to be shifted toward the northeast, with respect to the L-band map. If we compare the centroid and peak positions reported in Table~\ref{table:evn_cont}, we obtain that the differences in right ascension and declination are 4.8$\pm$0.3\,mas and 4.7$\pm$0.6\,mas, respectively, for component NE, while for the most compact source SW we get $\Delta {\rm RA} = 2.1 \pm 0.3$\,mas and $\Delta {\rm Dec} = 3.2 \pm 0.6$\,mas. Therefore, we find a remarkable shift of 6.7$\pm$0.8\,mas and 3.8$\pm$0.8\,mas, for components NE and SW, respectively. Using the positions derived from the 5\,GHz map convolved with the 1.7\,GHz beam, the value of the shift does not change significantly (6.6$\pm$0.8\,mas and 4.5$\pm$0.8\,mas, for components NE and SW, respectively).   
%Gli errori sullo shift sono la media quadratica degli errori assoluti sulle posizioni nelle due bande: sigma_lx=(sigma_l^2+sigma_c^2)^1/2 (vedi Hagiwara et al. 2001, paper su NGC5793).
%In questo caso sigma_l=sigma_c=0.6 mas

\begin{figure*}
\includegraphics[scale=0.45]{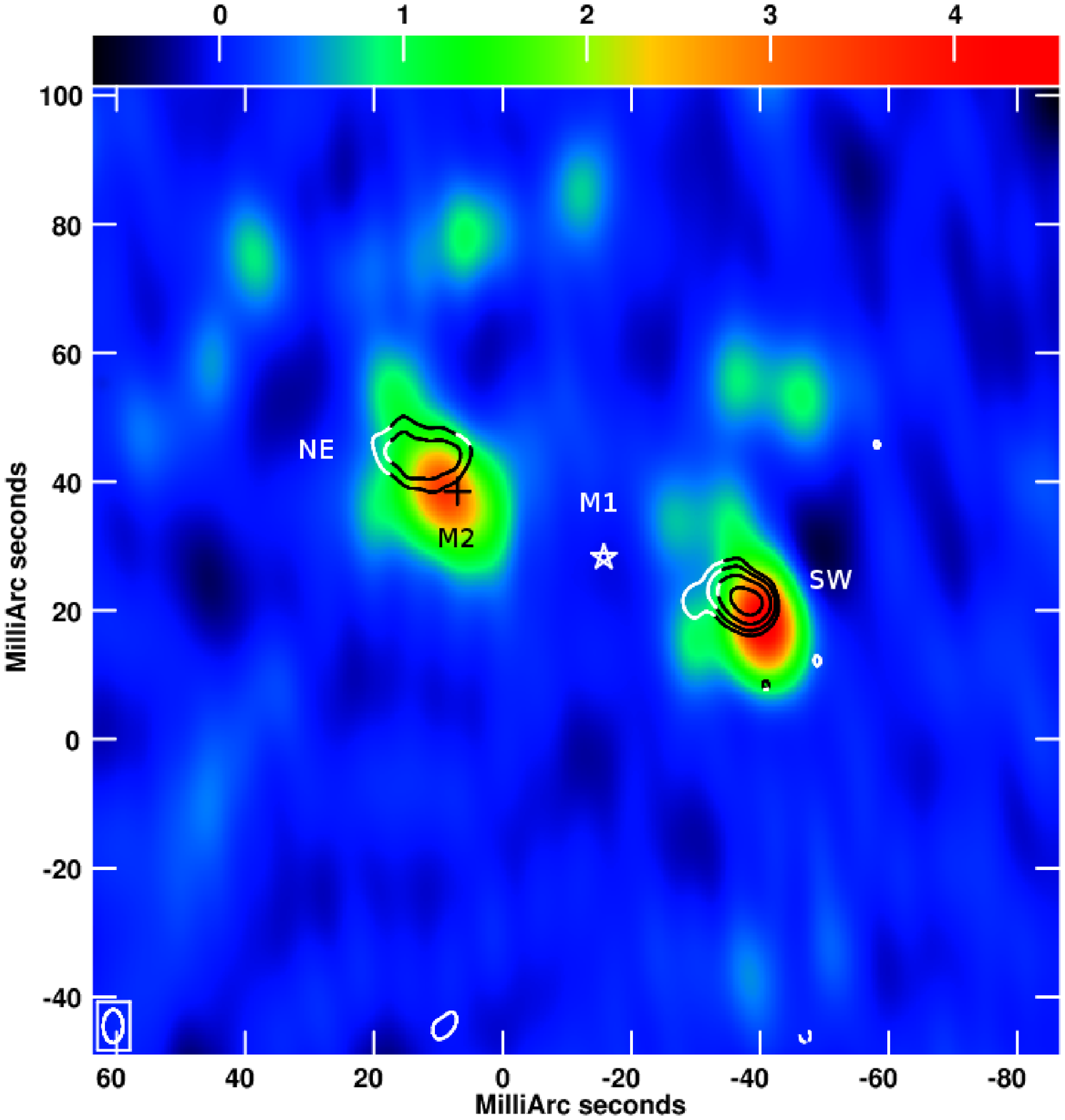}
\includegraphics[scale=0.45]{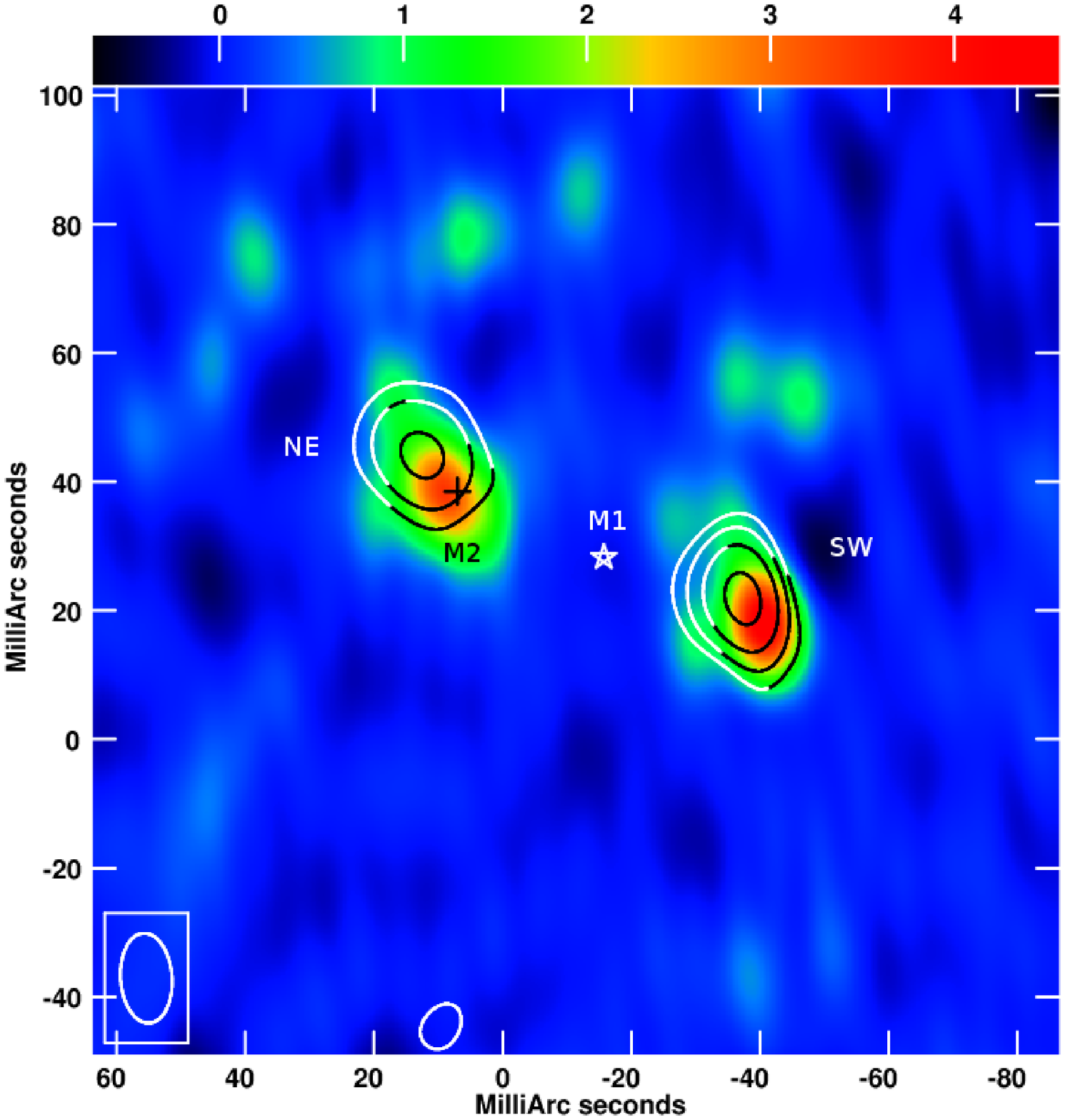}
\caption{Radio continuum emission in the nucleus of IRAS15480. The color scale represents emission at L-band, ranging from -0.7 to 4.6\,mJy/beam, while the overlaid contours delineate the C-band emission with the original restored beam ({\it left}) and convolved with the L-band beam ({\it right}). Contour levels are -1, 1, 2, 4, 8, 16, 32, 64 $\times$ the 5$\sigma$ noise level (0.2 and 0.45\,mJy/beam, for the original and convolved 5\,GHz emission, respectively). The positions of the water maser spots detected with the VLBA are also indicated. The star and the cross mark the location of the narrow (M1) and the broad blueshifted line emission (M2), respectively \citep[for details see][]{castangia2016}.} 
\label{fig:shift}
\end{figure*}

In principle this misalignment might be due to either a real astrophysical phenomenon or to a phase error during calibration.
Variation in the optical depth (e.\,g., synchrotron self-absorption) might cause the position of the source to change with frequency. There are two different scenarios:
\begin{enumerate}
\item the base of the jets (commonly called the ``core'') is not visible in {\bf the} EVN maps and its position is indicated by the M1 maser position (Fig~\ref{fig:shift}). In this case, we expect the C-band components to be both closer to the M1 position, with respect to the L-band ones. This is in contrast with what we observe;
\item component NE represents the radio core. In this scenario, the misalignment might be due to the ``core shift'' effect, a well-known frequency-dependent shift of the absolute position of the core. However, measured core shifts are usually much smaller than the one we found \citep[e.\,g.,][]{kovalev08}. Furthermore, the steep spectral index of component NE is not consistent with the hypothesis that this source is the core (see Sect~\ref{sect:cont_ne}). Finally, the shift is observed in both components toward the same direction. This fact requires the presence of two nuclei to be explained in terms of a core shift effect.
\end{enumerate}
All these arguments make the possibility that the misalignment is due to a real astrophysical phenomenon extremely unlikely.

\begin{figure*}
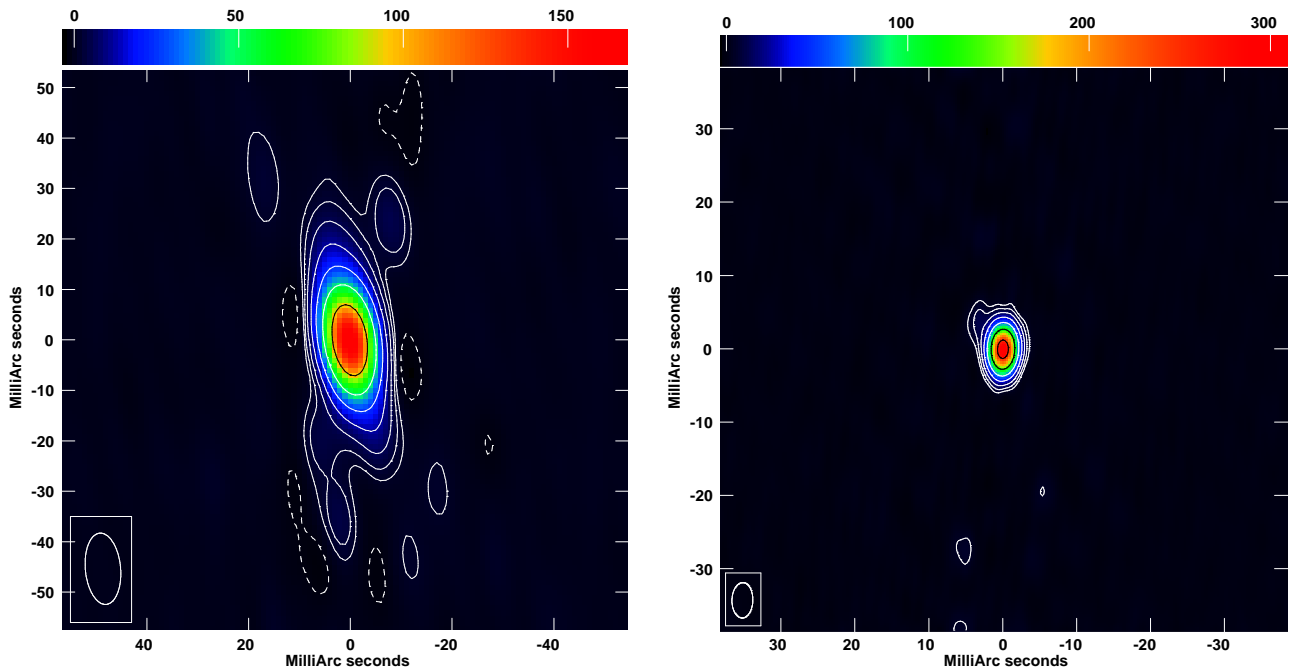

\includegraphics[scale=0.45]{castangia2019_fig9.eps}
\includegraphics[scale=0.45]{castangia2019_fig10.eps}
\caption{Images of the phase reference source J1555-0326 at 1.7\,GHz ({\it left}) and 5\,GHz ({\it right}). Contour levels are -1, 1, 2, 4, 8, 16, 32, 64 $\times$ the 5$\sigma$ noise level (0.3 and 0.8\,mJy/beam, for the 1.7 and 5\,GHz maps, respectively).} 
\label{fig:cal}
\end{figure*}

Potential phase errors include the dispersive refraction of the signal by the ionosphere and the presence of a resolved structure in the phase calibrator. The former effect, if not properly corrected, may produce a significant phase error at low frequency (e.g. at 1.7\,GHz). We used GPS models of the electron content of the ionosphere to correct for this dispersive delay. In order to inspect the image quality, we produced self-calibrated images of the phase reference source J1555-0326 at 1.7 and 5\,GHz (Fig.~\ref{fig:cal}; left and right panel, respectively). The 1.7\,GHz map shows a slightly resolved source (P.A.$\sim$20\degr) with a peak flux density $S_{1.7}\sim$170\,mJy. The morphology is consistent with the 2.3\,GHz map of J1555-0326 that can be found in the VLBA calibrator database\footnote{http://www.vlba.nrao.edu/astro/calib/}. However, despite self-calibration, the map looks somewhat dirty and the peak flux density is half of that measured at 2.3\,GHz (320\,mJy). Although the poor quality of the image is likely due to the highly irregular beam of the EVN observation and the low flux density can be explained by the lower frequency and the longer baselines of our EVN observations, nevertheless, we cannot exclude that these effects are due to residual atmospheric phase fluctuations. In addition, the direction-dependent propagation through the ionosphere introduces systematic position errors that are difficult to predict and correct, which can be of the order of a few mas \citep{rioja2017}. 

Inspection of the 5\,GHz image of J1555-0326 (Fig.~\ref{fig:cal}; right panel), instead, reveals an extension toward the north-east. The peak flux density is $S_{5}\sim$310\,mJy, in agreement with the values reported in the VLBA calibrator database. The north-east elongation is well visible also in the 2.3 and 8.3\,GHz maps of the phase calibrator and might cause the phase reference's position to change with frequency. In order to check this hypothesis, we forced the task FRING to ignore the extended structure when searching for the phase solutions. However, this work around did not resolve the problem. The misalignment between the position of the two continuum sources with frequency was still present. Therefore, we conclude that probably the resolved structure of J1555-0326 is not enough extended to cause a sensible position shift. The latter is more likely to be due to an uncompensated phase error in the L-band map. We then decided to take the C-band image as the ``correct'' one and referred the maser positions to this map. Subsequently, we shifted the L-band image by the number of pixels necessary to make the centroid of source SW to coincide at the two frequencies.
%We verified that doing the opposite, i.e. considering the L-band map as the ``correct'' one, does not affect the final results. Indeed, the position of the maser feature M2 is still coincident with the continuum source NE and with the steep spectrum region. 

\end{document}